# A human brain atlas of $\chi$-separation for normative iron and myelin distributions

Kyeongseon Min[1], Beomseok Sohn[2], Woo Jung Kim[3,4], Chae Jung Park[5], Soohwa Song[6], Dong Hoon Shin[6], Kyung Won Chang[7], Na-Young Shin[8], Minjun Kim[1], Hyeong-Geol Shin[9,10], Phil Hyu Lee[11], Jongho Lee[1,*]

[1]Department of Electrical and Computer Engineering, Seoul National University, Seoul, Republic of Korea
[2]Department of Radiology, Samsung Medical Center, Sungkyunkwan University School of Medicine, Seoul, Republic of Korea
[3]Institute of Behavioral Sciences in Medicine, Yonsei University College of Medicine, Seoul, Republic of Korea.
[4]Department of Psychiatry, Yongin Severance Hospital, Yonsei University College of Medicine, Yongin, Republic of Korea.
[5]Department of Radiology, Yongin Severance Hospital, Yonsei University College of Medicine, Yongin, Republic of Korea
[6]Heuron Co., Ltd, Republic of Korea
[7]Department of Neurosurgery, Severance Hospital, Yonsei University College of Medicine, Seoul, Republic of Korea
[8]Department of Radiology, Severance Hospital, Yonsei University College of Medicine, Seoul, Republic of Korea
[9]Department of Radiology, Johns Hopkins University School of Medicine, Baltimore, MD, USA
[10]F.M. Kirby Research Center for Functional Brain Imaging, Kennedy Krieger Institute, Baltimore, MD, USA
[11]Department of Neurology, Severance Hospital, Yonsei University College of Medicine, Seoul, Republic of Korea
*Correspondence: jonghoyi@snu.ac.kr

**Abstract**

Iron and myelin are primary susceptibility sources in the human brain. These substances are essential for healthy brain, and their abnormalities are often related to various neurological disorders. Recently, an advanced susceptibility mapping technique, which is referred to as $\chi$-separation (pronounced as "chi"-separation), has been proposed, successfully disentangling paramagnetic iron from diamagnetic myelin. This method opened a potential for generating high resolution iron and myelin maps in the brain. Utilizing this technique, this study constructs a normative $\chi$-separation atlas from 106 healthy human brains. The resulting atlas provides detailed anatomical structures associated with the distributions of iron and myelin, clearly delineating subcortical nuclei, thalamic nuclei, and white matter fiber bundles. Additionally, susceptibility values in a number of regions of interest are reported along with age-dependent changes. This atlas may have direct applications such as localization of subcortical structures for deep brain stimulation or high-intensity focused ultrasound and also serve as a valuable resource for future research.

# 1. Introduction

In the human brain, iron and myelin are essential components for normal brain functions[1]. Alterations in iron deposition or demyelination are often observed in neurodegenerative diseases such as multiple sclerosis, Parkinson's disease, and Alzheimer's disease[1], making iron and myelin as potential biomarkers for these diseases. Therefore, *in-vivo* imaging techniques capable of mapping iron and myelin distributions may have significant clinical value.

Magnetic resonance imaging (MRI) has been shown to be a potential tool that reveals iron and myelin distributions *in vivo*. For example, iron has been imaged using techniques such as relaxometry ($R_2$[2]; $R_2^*$ and $R_2'$[3]), susceptibility-weighted imaging (SWI)[4], and quantitative susceptibility map (QSM)[5]. Similarly, myelin has been visualized using magnetization transfer[6], $R_1$ maps[7], myelin water imaging (MWI)[8], and QSM. Among these techniques, QSM is shown to be sensitive to both iron and myelin and provides quantitative measurements[9-12].

Because QSM maps reflect the superposition of magnetic susceptibility sources, however, QSM encounters a limitation in distinguishing the contribution of iron and myelin. In particular, when iron and myelin are co-localized in a voxel, they result in the cancellation of paramagnetic iron and diamagnetic myelin contribution. To overcome this limitation, a susceptibility source separation method, which is referred to as $\chi$-separation, was recently proposed[13]. This method jointly utilizes phase and $R_2'$ (or $R_2^*$) to provide paramagnetic susceptibility ($\chi_{para}$) and diamagnetic susceptibility ($\chi_{dia}$) maps. Studies have demonstrated strong correlations between $\chi_{para}$ and iron, and between $\chi_{dia}$ and myelin, suggesting potential use of the two maps as surrogate biomarkers for iron and myelin[14-16]. Furthermore, this approach has been suggested to provide potentially meaningful information when evaluating neurological disorders such as multiple sclerosis and neuromyelitis optica spectrum disorder[13,17-20]. Alternative methodologies and technical developments have been continued, expanding future applications of the susceptibility source separation field[21-23].

In neuroimaging, a brain atlas serves as a common reference, facilitating comprehensive analysis of brain images acquired from diverse population and modalities, and has been widely utilized for investigating stereotactic anatomical features, conducting longitudinal studies, enabling automatic parcellation, and identifying neuropathological alterations[24]. With a novel MR image contrast, a corresponding atlas has been created to support and promote its utilization in research. Specifically, for iron and myelin imaging, QSM atlases[25-27], and MWI atlases[28,29] have been developed and utilized in various applications[26,29]. Recently, a susceptibility source separation atlas has also become available[30], which employed APART-QSM technique[23] to separate susceptibility sources.

In this study, our objective is to establish a normative $\chi$-separation atlas of the healthy human brain. Our atlas aims to visualize detailed anatomical structures related to iron and myelin distributions, and provide normal values of paramagnetic and diamagnetic susceptibility in anatomical regions.

The atlas may be useful in clinical applications for deep brain stimulation (DBS) or high-intensity focused ultrasound (HIFU) by localizing subcortical nuclei. Additionally, the atlas may help us to identify susceptibility alterations associated with neurodegenerative diseases. The $\chi$-separation atlas, along with regions of interest (ROI) labels, regional susceptibility values, and demographic data are made publicly available (https://github.com/SNU-LIST/chi-separation-atlas).

# 2. Methods

## 2.1. Subjects

A total of 116 healthy human volunteers aged 27–85 years (interquartile range: 47.5–76.5, median: 65, mean: 62.1 ± 15.8; 37 males and 79 females) were recruited from two centers (Yonsei Severance Hospital (YSSH) and Yongin Severance Hospital (YGSH)). The inclusion criteria are detailed in Supplementary Information. All subjects provided written informed consent. The study was reviewed and approved by the Institutional Review Board. After reviewing datasets, nine subjects were excluded due to severe calcification in globus pallidus, and one subject was excluded due to motion artifacts. Consequently, the study population was reduced to 106 volunteers aged 27–85 years (interquartile range: 46–75, median: 64, mean: 60.8 ± 15.8; 34 males and 72 females).

*2.2. MRI acquisition*

The volunteers were scanned using one of two 3 T MRI scanners (Ingenia CX in YSSH or Ingenia Elition X in YGSH, Philips Healthcare) with 32-channel receiver head coils. $T_1$-weighted and multi-echo GRE images were acquired for atlas construction. For the $T_1$-weighted images, a 3D magnetization-prepared rapid gradient echo (MPRAGE) sequence was utilized with acquisition parameters summarized in Supplementary Information. For the 3D multi-echo GRE, images were acquired using a monopolar gradient readout with the following scan parameters: TR = 33 ms, TE = 5.25, 11.08, 16.91, 22.74, and 28.57 ms, flip angle = 15°, resolution = 1×1×1 mm$^3$, FOV = 256×256×144 mm$^3$, pixel bandwidth = 228 Hz, and SENSE acceleration factor = 2. The imaging orientations of both sequences were aligned with the anterior commissure-posterior commissure line.

*2.3. χ-separation atlas construction*

The workflow for the χ-separation atlas is depicted in Fig. 1. First, the $T_1$-weighted image underwent $B_1$ bias field correction using a nonparametric nonuniform normalization algorithm[31] (*N4BiasFieldCorrection*, ANTs, https://github.com/ANTsX/ANTs). Subsequently, a brain mask was obtained by applying a template-based brain extraction algorithm (*antsBrainExtraction.sh*, ANTs) to the bias field-corrected $T_1$-weighted image.

The magnitude images of all echoes in the multi-echo GRE data were averaged and corrected for a $B_1$ bias field (*N4BiasFieldCorrection*, ANTs), generating a combined GRE image. Then, the $T_1$-weighted image was registered to this image using an affine transformation with six degrees of freedom (*antsRegistration*, ANTs). The brain mask from the $T_1$-weighted image was also transformed to the multi-echo GRE space using the affine transformation.

To generate a χ-separation map, a processing pipeline established in the χ-separation toolbox (https://github.com/SNU-LIST/chi-separation) was employed. The χ-separation model and its implementation were reported previously[13,32]. Briefly, it is based on a biophysical model that describes the contributions of paramagnetic susceptibility ($\chi_{para}$) and diamagnetic susceptibility ($\chi_{dia}$) to the magnetic field perturbation ($\Delta f$) and the reversible transverse relaxation rate ($R_2' = R_2^* - R_2$). The relationship is expressed as follows[13]:

$$\Delta f = D_f * (\chi_{para} + \chi_{dia}) \quad (1)$$

$$R_2' = D_{r,para} |\chi_{para}| + D_{r,dia} |\chi_{dia}| \quad (2)$$

where $D_f$ is the dipole kernel, and $D_{r,para}$ and $D_{r,dia}$ are the relaxometric constants relating $R_2'$ to the paramagnetic and diamagnetic susceptibility, respectively. In our study, a neural network implementation of χ-separation which is called χ-sepnet-$R_2^*$ was employed[32] (https://github.com/SNU-LIST/chi-separation). This neural network was trained using multi-head orientation data as labels to provide streaking artifact-free $\chi_{para}$ and $\chi_{dia}$ maps[33]. Additionally, the network was trained to use $R_2^*$ instead of $R_2'$ as the input data, allowing the creation of $\chi_{para}$ and $\chi_{dia}$ without $R_2$. The outcome of this network has demonstrated a good agreement with that using $R_2'$ input both in healthy volunteers[32] and multiple sclerosis patients[34] at the cost of slight degradation in accuracy.

The inputs for χ-sepnet-$R_2^*$, which were a tissue field map ($\Delta f$) and an $R_2^*$ map, were obtained from the multi-echo GRE data via the following steps. First, the phase maps from the multi-echo GRE data were unwrapped using the Laplacian-based unwrapping algorithm[35] (*MRPhaseUnwrap*, STI Suite, https://people.eecs.berkeley.edu/~chunlei.liu/software.html). The unwrapped phase maps were averaged using SNR-weighted mean[36] to produce a combined phase map. Then, the brain mask was applied. The result was processed to remove background field by applying V-SHARP[37] (STI Suite), generating a tissue field map ($\Delta f$). An $R_2^*$ map was generated by fitting a mono-exponential decay function to the multi-echo magnitude signals using a nonlinear least square solver (*lsqnonlin*, MATLAB 2016b). Finally, $\Delta f$ and $R_2^*$ maps were applied to χ-sepnet-$R_2^{*}$[32] (*chi_sepnet_general*, *chi*-separation toolbox), creating $\chi_{para}$ and $\chi_{dia}$ maps. These two maps were summed to produce a QSM map.

This QSM map was linearly combined with the $T_1$-weighted image, creating a hybrid image in order to achieve accurate registration of subcortical structures[26,38,39]. Before the combination, the histogram of a $T_1$-weighted image was normalized by a piecewise linear transformation that maps the deciles of the image to the averaged deciles of all $T_1$-weighted images[40]. The intensity-normalized $T_1$-weighted image was scaled to the range of 0–255. Then the hybrid image was calculated as follows[39]:

$$[\text{Hybrid image}] = [\text{Corrected } T_1\text{-weighted image}] - 0.8 \text{ (ppb}^{-1}) \times [\text{QSM map (in ppb)}]. \quad (3)$$

This hybrid image was registered to the hybrid image atlas in the MNI space from MuSus-100[39], using a symmetric diffeomorphic image normalization algorithm[41] (*antsRegistration*, ANTs). Subsequently, the $\chi_{para}$ and $\chi_{dia}$ maps were transformed to the MNI space using the same deformation field. The $\chi_{para}$ and $\chi_{dia}$ maps in the MNI space from the 106 subjects were averaged, creating $\chi_{para}$ and $\chi_{dia}$ atlases. Similarly, QSM, hybrid, and $T_1$-weighted image atlases were created by averaging these images in the MNI space. To evaluate the subject-dependent variability of the atlases, a voxel-wise relative standard deviation (*rSD*) was calculated, creating the *rSD* maps of $\chi_{para}$ and $\chi_{dia}$.

$$[rSD \text{ of a voxel}] (\%) = [\text{Standard deviation of all subjects}] / | [\text{Mean of all subjects}] | \times 100. \quad (4)$$

*2.4. Atlas analysis*

To explore the characteristics of the $\chi$-separation atlas, regions of interest (ROIs) were selected in subcortical nuclei, thalamic nuclei, and white matter fiber bundles. For the subcortical and thalamic nuclei, parcellation labels from MuSus-100 atlas were employed. Thalamic nuclei ROIs in MuSus-100 were manually refined into four ROIs: medial thalamic nuclei, lateral thalamic nuclei, pulvinar, and whole thalamus. For white matter, the $T_1$-weighted image atlas with white matter labels from ICBM-DTI-81[42] was registered to our $T_1$-weighted image atlas. Then, the white matter labels were eroded with a radius of 1 mm to reduce registration errors, defining white matter ROIs. Out of twenty-eight ROIs, the largest thirteen ROIs were selected, discarding small ROIs with large registration errors. Additionally, a whole white matter mask, generated from the hybrid images via tissue segmentation[43], was included as white matter ROI. Consequently, eight subcortical nuclei, four thalamic nuclei, and fourteen white matter ROIs were employed (Fig. S1). Prior to the ROI analysis, the intersection of the subcortical nuclei, thalamic nuclei, and white matter ROIs was excluded to ensure that they were mutually exclusive. Then, median $\chi_{para}$, $\chi_{dia}$, and QSM values were extracted from the ROIs. Median was used instead of mean to enhance statistical robustness against outlier voxels such as vessels and calcification within the ROIs. The median values were averaged across subjects to report the population means and standard deviations.

The population-averaged $\chi_{para}$ and QSM values in the five subcortical nucleus ROIs were linearly regressed with respect to the non-heme iron content from literature[44] to demonstrate the well-known linear relationship[9,25,45]. The $R^2$ value, regression line and its 95% confidence interval were calculated and the significance of the regression coefficients was assessed using *t*-tests.

To examine the relationship of $\chi_{dia}$ and myelin content, a myelin water fraction (MWF) atlas[28] was employed as a surrogate marker for myelin content. Before the analysis, the $T_1$-weighted image atlas (ICBM-152) accompanied with the MWF atlas were registered to our $T_1$-weighted image atlas, and then the MWF atlas was transformed using the same deformation field. The $\chi_{dia}$ values in the white matter ROIs were linearly regressed with respect to the MWF values in the same ROIs. The $R^2$ value, regression line, and its 95% confidence interval were calculated and the significance of the regression coefficients was assessed. To account for the errors in both $\chi_{dia}$ and MWF values, Deming regression was performed to fit the observations to a regression line.

Lastly, age-dependent iron and myelin changes were explored for each ROI. The individual median values of $\chi_{para}$ and $\chi_{dia}$ in each ROI were linearly regressed against age. The $R^2$ value, the regression line, and its 95% confidence interval were calculated and the significance of the regression coefficients was assessed using *t*-tests. The significance level was Bonferroni-corrected, considering the total number of ROIs, which was 26.

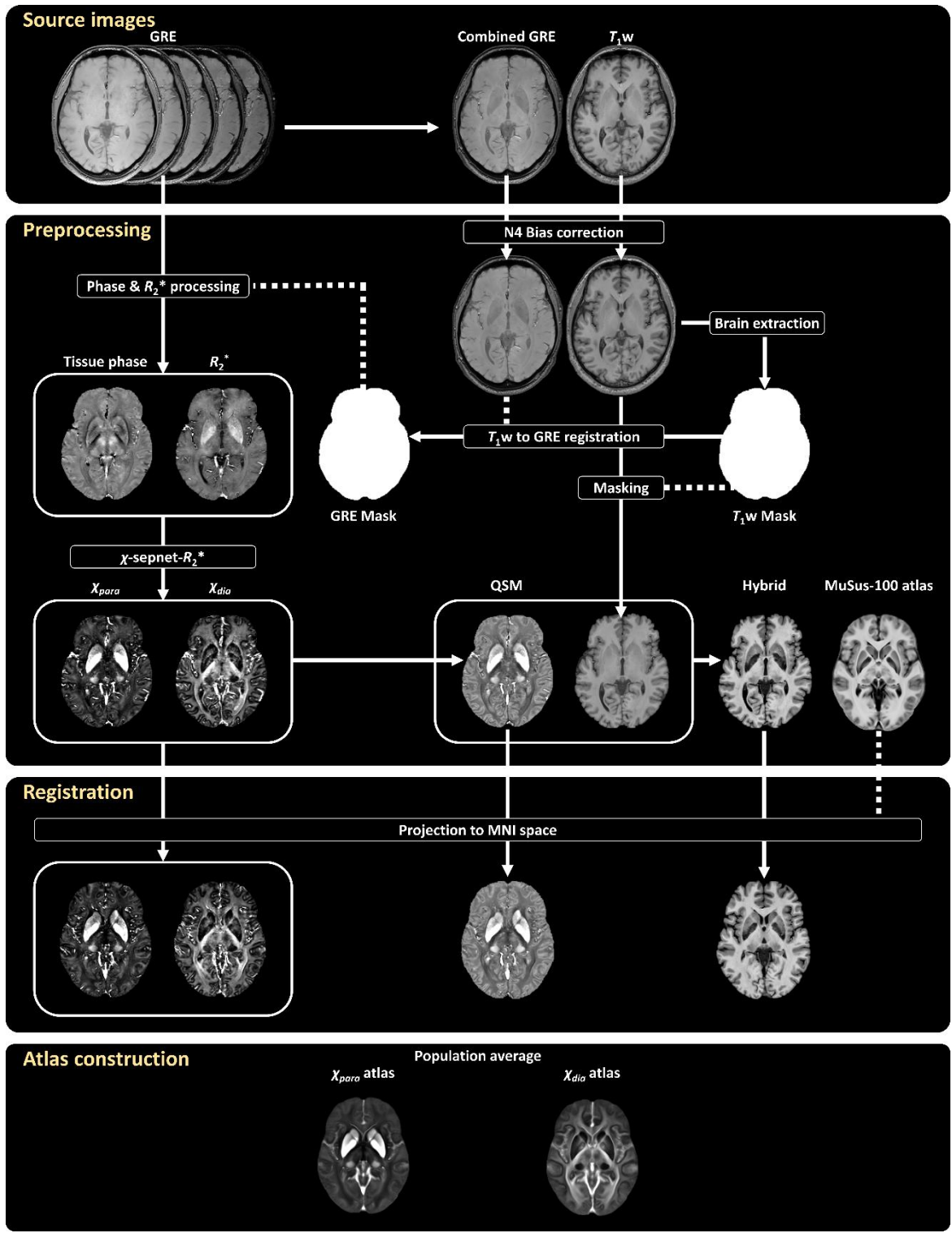

**Fig. 1.** Schematic flowchart illustrating the processing pipeline for generating the $\chi_{para}$ and $\chi_{dia}$ atlases. The hybrid images were created by a linear combination of the QSM and $T_1$-weighted images to improve registration in the deep gray matter area.

## 3. Results

### 3.1. $\chi_{para}$ and $\chi_{dia}$ atlases

Figure 2 shows representative slices in the axial, coronal and sagittal planes of the population-averaged $\chi_{para}$ and $\chi_{dia}$ atlases and corresponding *rSD* maps. The $\chi_{para}$ atlas highlights regions rich in iron, such as caudate nucleus, putamen, globus pallidus, pulvinar, subthalamic nucleus, red nucleus, substantia nigra, and ventral pallidum, which exhibit high $\chi_{para}$ values. On the other hands, white matter structures reveal high $|\chi_{dia}|$ values in corpus callosum, external capsule, internal capsule, optic radiation, optic tract, cingulum, fornix, and cerebellar peduncle. It should be noted that artifacts, mostly from large vessels, exist in both atlases (see Discussion and Fig. S2). The *rSD* maps have high values in regions with low susceptibility and/or regions close to large vessels and ventricles.

In Fig. 3, ten axial slices of the $\chi_{para}$ and $\chi_{dia}$ atlases that contain distinctive susceptibility distributions are displayed. Regions not covered in Fig. 2, such as the dentate nucleus (DN, $z = -38$ mm), corticospinal tract (CST, $z = -30$ mm), pontine crossing tract (PCT, $z = -30$ mm), and hand knob of primary motor cortex (M1, $z = +55$ mm), are visible in the axial slices. Overall, the $\chi_{para}$ and $\chi_{dia}$ atlases provide detailed anatomical information that reflects well-known distributions of iron and myelin in the brain.

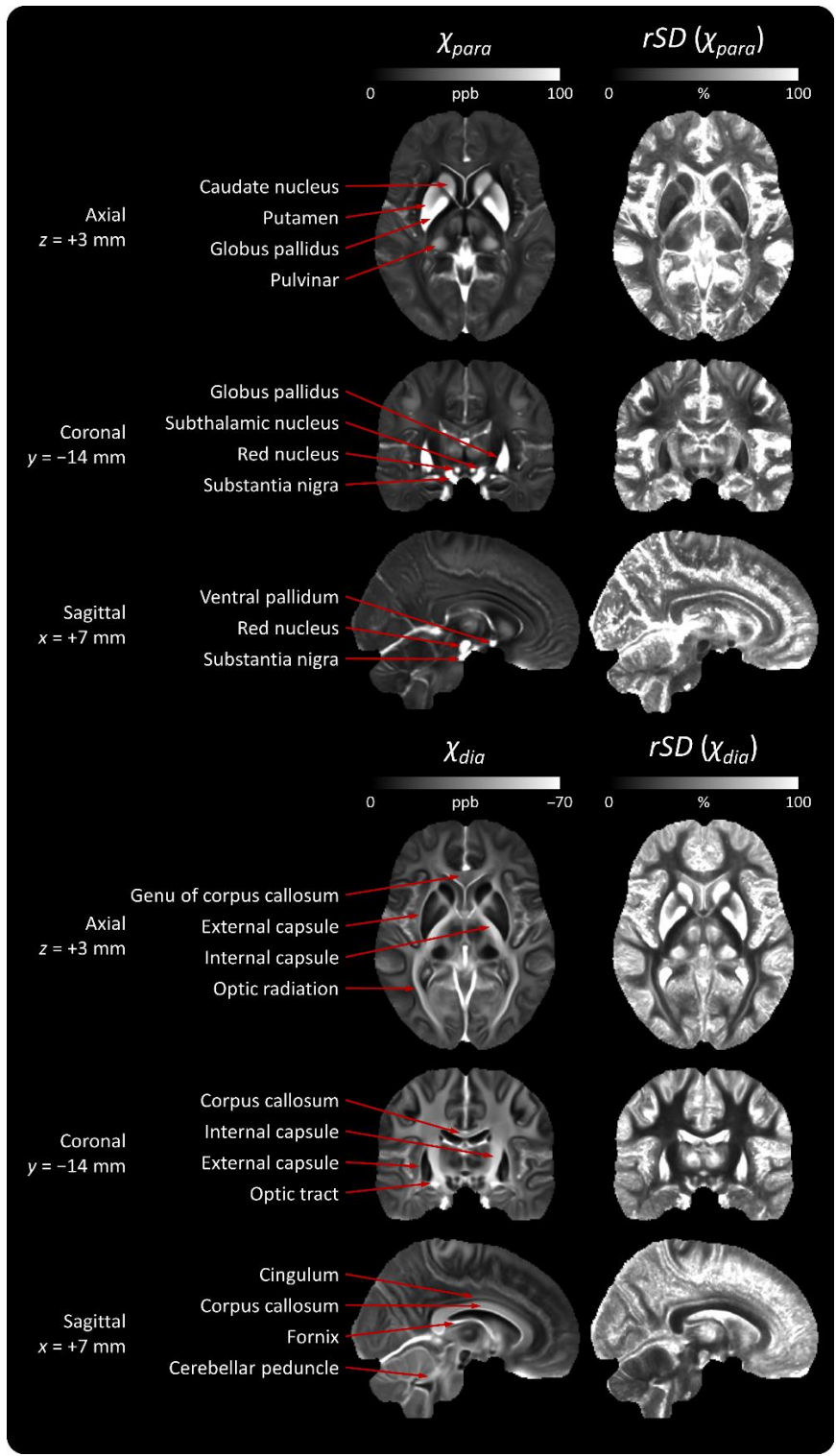

**Fig. 2.** Representative axial, coronal, and sagittal slices of the $\chi_{para}$ and $\chi_{dia}$ atlases and *rSD* maps.

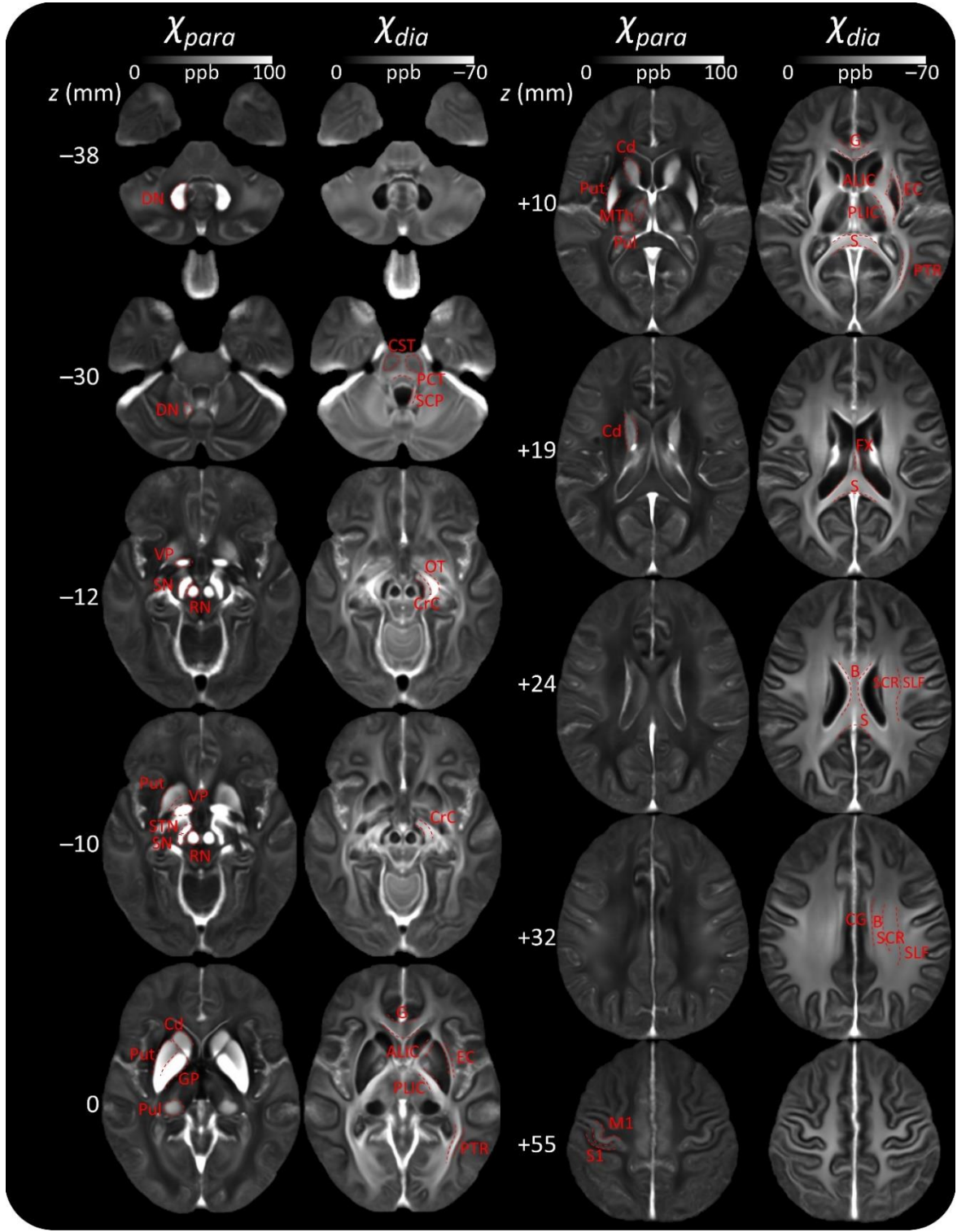

**Fig. 3.** Ten axial slices of the $\chi_{para}$ and $\chi_{dia}$ atlases. Anatomical structures are marked with red dotted lines. [DN: dentate nucleus, SN: substantia nucleus, VP: ventral pallidum, RN: red nucleus, Put: Putamen, STN: subthalamic nucleus, Cd: caudate nucleus; GP: globus pallidus, Pul: pulvinar, MTh: medial thalamic nuclei group, PCT: pontine crossing tract, CST: corticospinal tract, SCP: superior cerebellar peduncle, OT: optic tract, CrC: crus cerebri, ALIC: anterior limb of internal capsule, PLIC: posterior limb of internal capsule, EC: external capsule, PTR: posterior thalamic radiation, G: genu of corpus callosum, B: body of corpus callosum, S: splenium of corpus callosum, FX: fornix, CG: cingulum, SCR: superior corona radiata, SLF: superior longitudinal fasciculus, M1: primary motor cortex, S1: primary somatosensory cortex]

*3.2. Details of anatomical structures*

When basal ganglia and midbrain areas are examined, they reveal exquisite details of anatomy (Fig. 4). The zoomed-in images of the yellow rectangles are presented next to histology images from the literatures (Perls' stain images from[46]; Heidenhain-Woelcke's stain images from[47]). The iron distribution in the stained images, which are stained in cyan (darker color means more iron), is qualitatively well-delineated in the $\chi_{para}$ maps, revealing fine structures like claustrum (Cl, upper row), anterior (A)/medial (M)/lateral (L) thalamic nuclei groups (middle row), and external (GPe) and internal (GPi) globus pallidus (bottom row). Some of these structures may not be easily identified in other MRI images. The $\chi_{dia}$ maps reflect myelin distribution in the myelin stained images (darker color means more myelin). For example, external capsule (EC, upper row), posterior limb of internal capsule (PLIC, middle row) and anterior limb of internal capsule (ALIC, lower row) demonstrate high concentrations of myelin while Cl, Put (all rows) and Cd (bottom row) are lacking myelin. In particular, lateral thalamic nucleus (L) possesses moderate concentrations of iron and myelin in the histology, which are also correctly-depicted in the $\chi$-separation maps, supporting successful separation of the para- and diamagnetic susceptibility sources (middle row). In contrast, the other thalamic nuclei (A, M, and Pul) show high concentrations of iron with no or little sign of myelin while PLIC suggests the opposite characteristics. All of these observations are well-matched to the $\chi$-separation results (middle row). Lastly, lateral medullary laminae (LML), which separates the putamen (Put) and external globus pallidus, and medial medullary laminae (MML), which separates the external and internal globus pallidus, are also observed (bottom row).

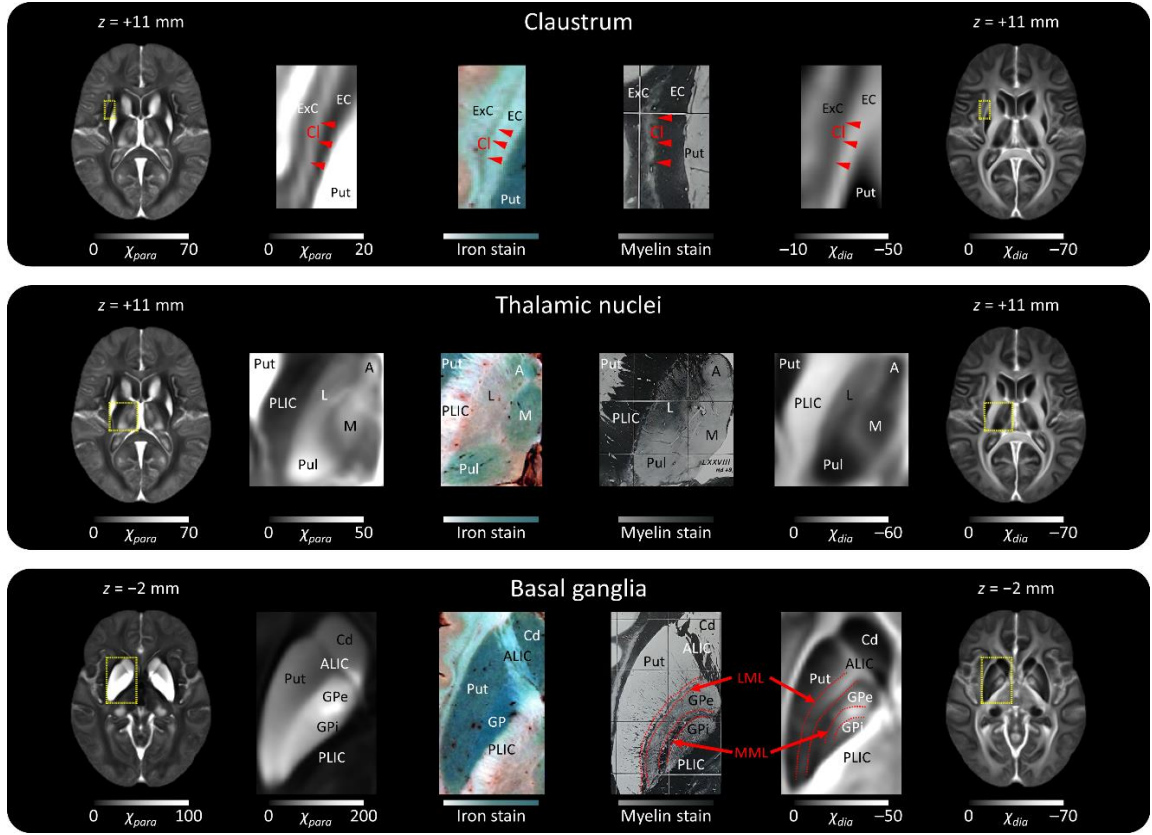

**Fig. 4.** Structures in the $\chi_{para}$ and $\chi_{dia}$ atlases presented along with the corresponding iron and myelin staining images. The zoomed-in images correspond to the regions indicated with yellow rectangles in the axial slices. The iron staining images are adapted and resized from Naidich et al.[46]. The myelin staining images are adapted and resized from Schaltenbrand and Wahren[47]. The claustrum (Cl), lateral (LML) and medial (MML) medullary lamina are marked with red arrows. [ExC: extreme capsule, EC: external capsule, Cl: claustrum, Put: Putamen, A: anterior thalamic nuclei group, M: medial thalamic nuclei group, L: lateral thalamic nuclei group, Pul: pulvinar, ALIC: anterior limb of internal capsule, PLIC: posterior limb of internal capsule, LML: lateral medullary lamina, MML: medial medullary lamina, Cd: caudate nucleus, GP: globus pallidus, GPe: external globus pallidus, GPi: internal globus pallidus]

*3.3. ROI analysis*

The means and standard deviations of $\chi_{para}$, $\chi_{dia}$, and QSM in the ROIs are plotted in Fig. 5 and summarized in Table S1. All the subcortical nuclei reveal high values of $\chi_{para}$ and much lower values of $\chi_{dia}$, indicating predominant concentration of iron in the regions. The ventral pallidum ROI exhibits the highest $\chi_{para}$ value of 144.3 ± 27.0 ppb, followed by globus pallidus (131.9 ± 10.4 ppb), substantia nigra (115.7 ± 14.4 ppb), subthalamic nucleus (112.1 ± 12.1 ppb), red nucleus (112.0 ± 14.8 ppb), putamen (77.1 ± 20.6 ppb), and pulvinar (52.1 ± 11.6 ppb). In thalamus, the lateral thalamic nuclei ROI reports comparable levels of $\chi_{para}$ (22.0 ± 5.2 ppb) and $\chi_{dia}$ (–22.4 ± 4.7 ppb), which are in good agreement with the iron and myelin staining images in Fig. 4, demonstrating successful separation of para- and diamagnetic susceptibility. On the other hand, medial thalamic nuclei and pulvinar ROIs yield higher concentration of $\chi_{para}$ (medial thalamic nuclei = 31.2 ± 7.6 ppb and pulvinar = 52.1 ± 11.6 ppb) than $\chi_{dia}$ (medial thalamic nuclei = –11.8 ± 5.9 ppb and pulvinar = –4.4 ± 3.2 ppb). These measurements also qualitatively agree with the corresponding areas in the iron and myelin staining images of Fig. 4. In the white matter ROIs, $|\chi_{dia}|$ values are higher than $\chi_{para}$ values although a few ROIs (e.g., superior longitudinal fasciculus) report moderate values of $\chi_{para}$. The posterior limb of internal capsule has the lowest $\chi_{dia}$ value of –52.2 ± 3.0 ppb (or the highest negative susceptibility). The second lowest $\chi_{dia}$ value is observed in the cerebral peduncle (–43.7 ± 4.1 ppb) followed by splenium of corpus callosum (–42.3 ± 3.1 ppb), anterior limb of internal capsule (–41.3 ± 4.0 ppb), and posterior thalamic radiation (–41.1 ± 4.0 ppb).

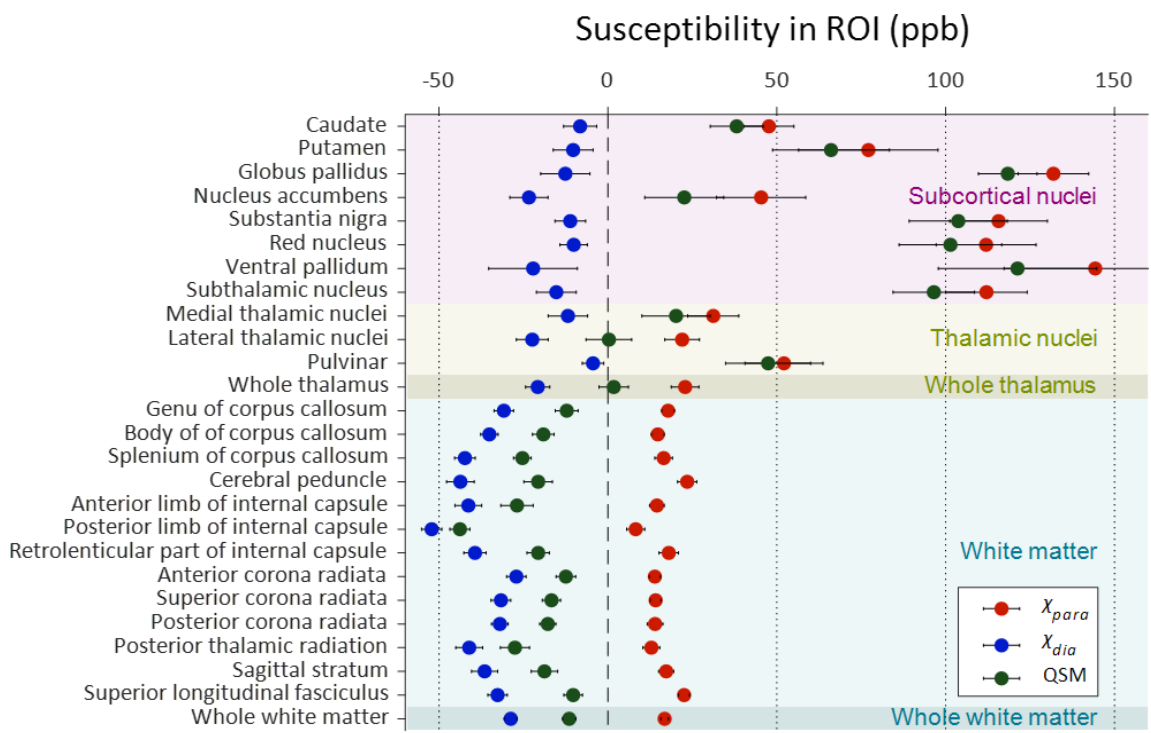

**Fig. 5.** The population averages of $\chi_{para}$, $\chi_{dia}$, and QSM medians extracted from 26 ROIs (number of subjects = 106). Red, blue, and green circles represent the $\chi_{para}$, $\chi_{dia}$, and QSM averages, respectively. The black line attached to the circle indicates the standard deviation value of each ROI.

When the $\chi_{para}$ ROI values are further inspected by plotting them against the postmortem iron content measurements[44], they report a clear linear trend ($R^2 = 0.99$, Fig. 6A). The linear regression line is:

$$\chi_{para} \text{ (ppb)} = 6.8 \times \text{iron content (mg/100 g)} - 15. \qquad (5)$$

While the linear coefficient is statistically significant ($p < 0.001$), the intercept is not ($p = 0.15$).

Likewise, when the QSM values are linearly regressed with the iron content, a similar linear trend is observed ($R^2 = 0.99$, Fig. 6B), consistent with the previous QSM studies[25,45]. The regression line is:

$$\text{QSM (ppb)} = 6.6 \times \text{iron content (mg/100 g)} - 22. \tag{6}$$

Here, both the linear coefficient ($p < 0.001$) and the intercept ($p < 0.05$) are statistically significant. This intercept value, which is only significant in the QSM result, demonstrates that a non-zero offset exists between the iron and QSM relationship. This offset, potentially from diamagnetic susceptibility sources, is removed in $\chi_{para}$ results, reporting a reduced offset (statistically nonsignificant), suggesting that $\chi_{para}$ measurements better reflects iron than those of QSM.

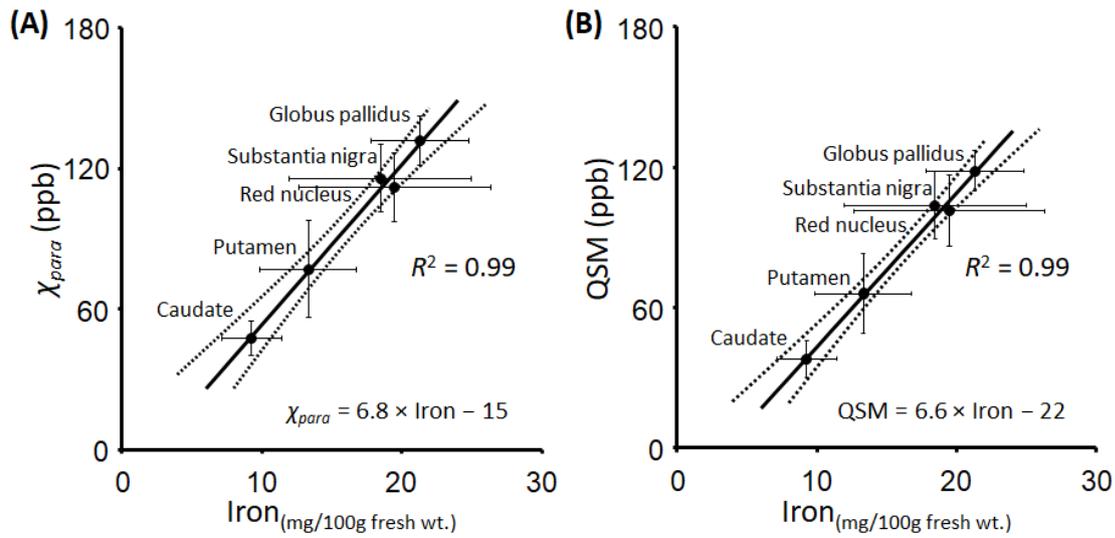

**Fig. 6.** Linear regression analysis between (A) $\chi_{para}$ or (B) QSM and iron content from the literature[44]. The regression line is drawn with a solid line along with dotted lines representing 95% confidence intervals.

The $\chi_{dia}$ atlas and MWF atlas are visually compared in Fig. 7A. Overall, the $\chi_{dia}$ and MWF atlases show high values in myelin-rich structures such as posterior limb of internal capsule, posterior thalamic radiation, and splenium of corpus callosum, demonstrating a notable correspondence between the two atlases. Some areas (e.g. superior corona radiata), however, reveal differences potentially due to biases of the methods (e.g. fiber orientation effects). Figure 7B shows the results of linear regression analysis between the white matter ROI values of our $\chi_{dia}$ atlas and MWF atlas. The analysis shows a significant positive correlation ($R^2 = 0.71$). The linear regression line is:

$$|\chi_{dia}| \text{ (ppb)} = 3.6 \times \text{MWF (\%)} - 3.3. \tag{7}$$

The linear coefficient is statistically significant ($p < 0.05$), but the intercept is not ($p = 0.57$).

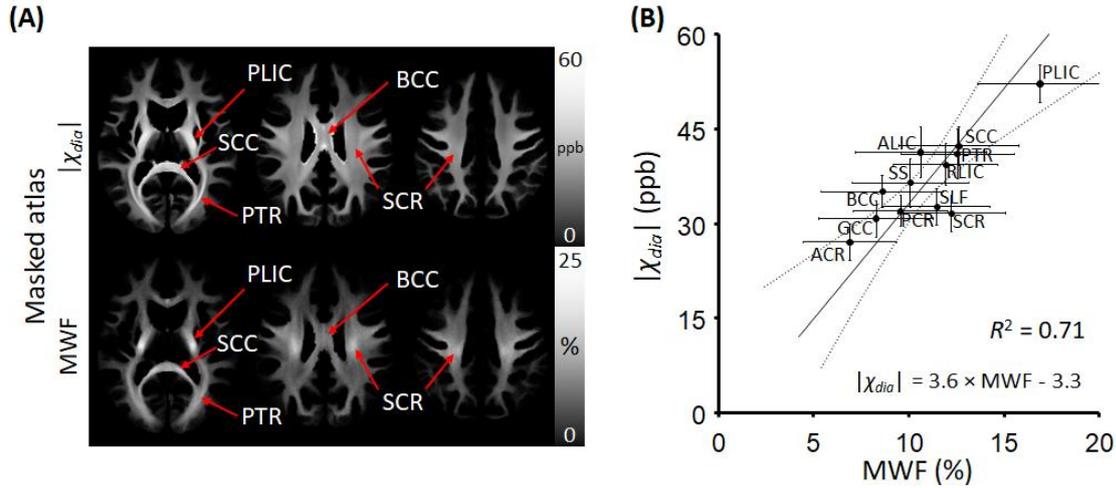

**Fig. 7.** The comparison between our $\chi_{dia}$ atlas and an MWF atlas from the literature[28]. (A) A visual comparison of three representative slices of the $\chi_{dia}$ atlas and MWF atlas. The atlases are masked with the whole white matter ROI. (B) Linear regression analysis of the white matter ROIs values. The regression line is drawn with a solid line along with dotted lines representing 95% confidence intervals. [GCC: genu of corpus callosum, BCC: body of corpus callosum, SCC: splenium of corpus callosum, ALIC: anterior limb of internal capsule, PLIC: posterior limb of internal capsule, RLIC: retrolenticular part of internal capsule, ACR: anterior corona radiata, SCR: superior corona radiata, PCR: posterior corona radiata, PTR: posterior thalamic radiation, SS: sagittal striatum, SLF: superior longitudinal fasciculus]

*3.4. Age-dependence of $\chi_{para}$ and $\chi_{dia}$*

The age-dependent susceptibility changes in the subcortical and thalamic nuclei ROIs are summarized in Fig. 8 (Table S2 for the statistics of linear regression). Putamen and nucleus accumbens (Fig. 8B, D) show significantly increasing trends in $\chi_{para}$ (putamen, $p = 1.6 \times 10^{-12}$; nucleus accumbens, $p = 1.8 \times 10^{-15}$; Bonferroni-corrected), whereas thalamus (Fig. 8L) shows a significantly decreasing trend (whole thalamus, $p = 1.7 \times 10^{-7}$; Bonferroni-corrected). Similar decreases are observed in medial thalamic nuclei ($p = 2.1 \times 10^{-4}$) and lateral thalamic nuclei ($p = 7.2 \times 10^{-7}$). The trends of putamen and thalamus are consistent with the age-dependent iron deposition reported in the literature[44]. Caudate, globus pallidus, substantia nigra, red nucleus, ventral pallidum, and subthalamic nucleus show slightly positive slopes with no statistically significance. For $|\chi_{dia}|$, statistically significant increases are observed in caudate, putamen, globus pallidus, nucleus accumbens, and ventral pallidum whereas no significant changes were observed in the other ROIs.

Figure 9 shows the age-dependent susceptibility analysis in the white matter ROIs (Table S2 for the statistics of linear regression). Overall, white matter ROIs show decreasing trends in $|\chi_{dia}|$, with the whole white matter ROI revealing a significant decrease over the age range ($p = 2.7 \times 10^{-3}$, Bonferroni-corrected). This trend agrees with the previous studies using myelin water imaging[48-50], suggesting a decrease of myelin with age. For $\chi_{para}$, statistically significant decreases are observed in a few white matter ROIs while the other ROIs show almost no change.

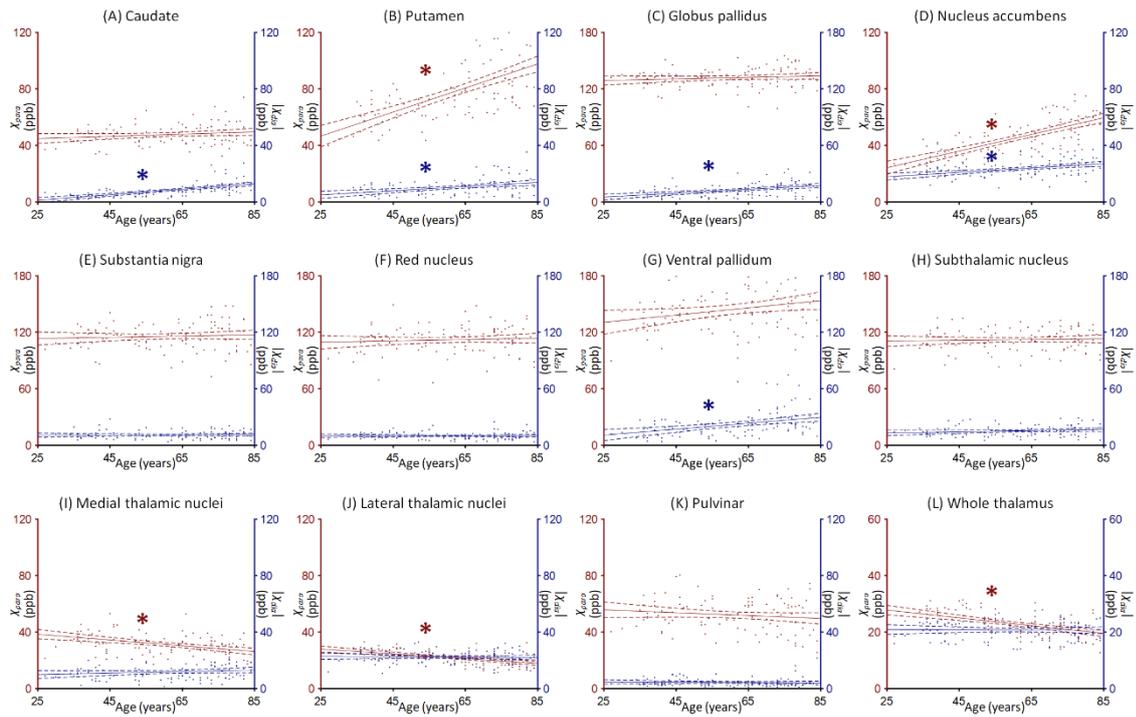

**Fig. 8.** The age-dependence of the $\chi_{para}$ and $\chi_{dia}$ values in the subcortical and thalamic nuclei ROIs: (A) Caudate, (B) putamen, (C) globus pallidus, (D) nucleus accumbens, (E) substantia nigra, (F) red nucleus, (G) ventral pallidum, (H) subthalamic nucleus, (I) medial thalamic nuclei, (J) lateral thalamic nuclei, (K) pulvinar, and (L) whole thalamus. The ordinate axes, data points, regression lines, and significance markers for $\chi_{para}$ and $\chi_{dia}$ values are colored with red and blue, respectively. The regression line is drawn with a solid line along with dotted lines representing 95% confidence intervals. The significance of the regression slope is marked with an asterisk (*, $p < 0.05$, Bonferroni-corrected).

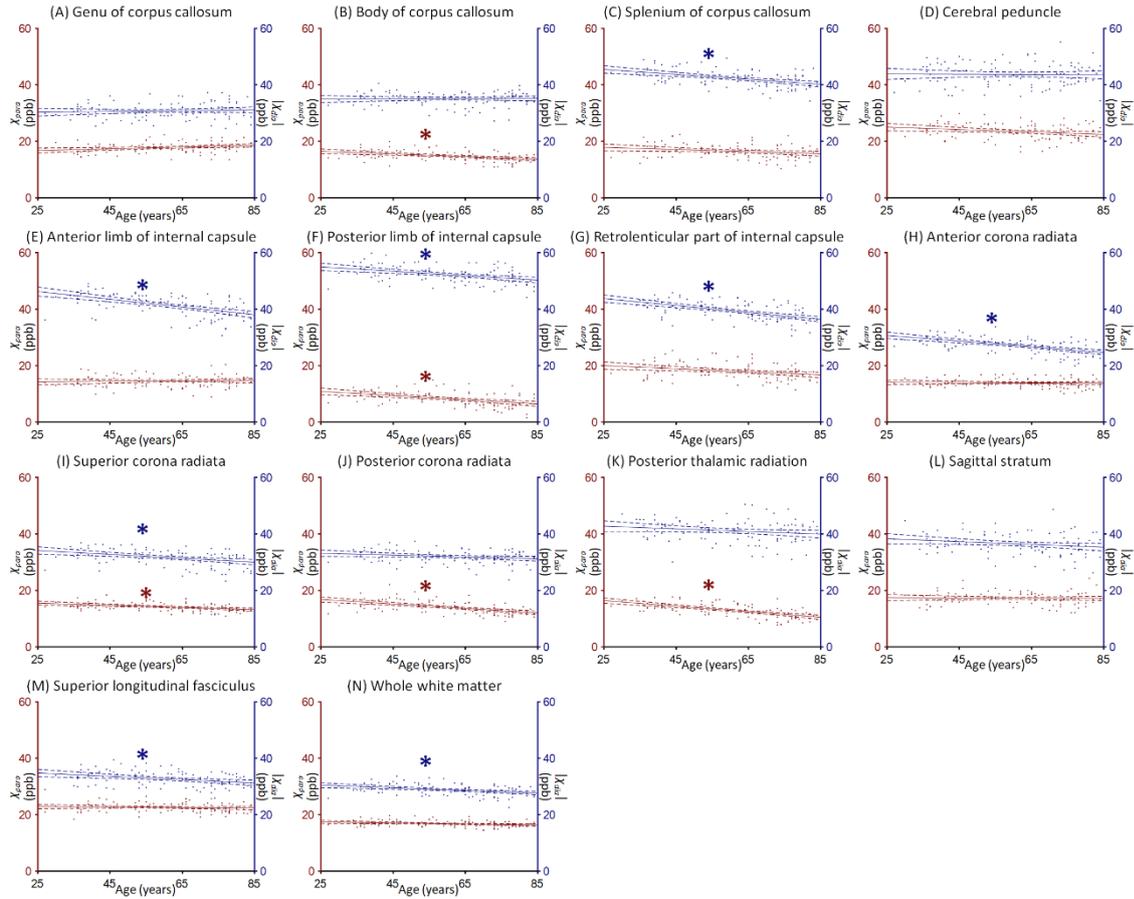

**Fig. 9.** The age-dependence of the $\chi_{para}$ and $\chi_{dia}$ values in the white matter ROIs: (A) Genu, (B) body, and (C) splenium of corpus callosum, (D) cerebral peduncle, (E) anterior limb, (F) posterior limb, and (G) retrolenticular part of internal capsule, (H) anterior, (I) superior, and (J) posterior corona radiata, (K) posterior thalamic radiation, (L) sagittal stratum, (M) superior longitudinal fasciculus, and (N) whole white matter. The ordinate axes, data points, regression lines, and significance markers for $\chi_{para}$ and $\chi_{dia}$ values are colored with red and blue, respectively. The regression line is drawn with a solid line along with dotted lines representing 95% confidence intervals. The significance of the regression slope is marked with an asterisk (*, $p < 0.05$, Bonferroni-corrected).

## 4. Discussion

In this study, we developed a $\chi$-separation atlas for healthy population. Our atlas provides distributions of $\chi_{para}$ and $\chi_{dia}$ in the brain, successfully revealing details of subcortical nuclei, thalamic nuclei, and white matter fiber bundles.

The atlas may contain errors that have been identified in the previous $\chi$-separation studies[13,23]. The most notable source of errors is vessels in the brain (Fig. S2). Flow artifacts can hinder the accurate evaluation of tissue phase and $R_2^*$ maps, leading to erroneous $\chi_{para}$ and $\chi_{dia}$ values. Additionally, vessels show large variability among subjects, creating unreliable regions with high *rSD*. Another major source of errors is anisotropy in the microstructure and susceptibility of myelin that may introduce fiber-orientation dependent variations in $\chi$[51,52] and $R_2^*$[53,54]. This may introduce potentially higher $|\chi_{dia}|$ in perpendicular fibers than those in parallel fibers. The assumption of a constant $D_r$ ($D_{r,para}$ and $D_{r,dia}$) across brain regions and among subjects may also introduce errors in the estimated susceptibility results, which will be a topic of continuing research[23,55].

The age distribution of our study spread over the range of 27–85 but the mean age of the population was 60.8 ± 15.8 years because the atlas is targeted toward population of neurodegenerative diseases. Therefore, caution should be exercised when applying our atlas to a young population (e.g., 20–30 years old). Considering the age-dependence nature of iron accumulation[44] and myelination[56], another atlas is expected for the range of 0–30 years old that covers different characteristics of iron and myelin changes.

In our analysis, a linear function was fitted to explore age-dependent changes whereas a nonlinear trend was reported for non-heme iron accumulation over age [44]. We believe the linear fit is reasonable considering the study population (27–85 years old with the mean age of 60.8 years) because rapid iron accumulation occurs in the first two decades after birth whereas slower accumulation happens after the age of 30[44]. Our results demonstrated significant iron accumulation over age in putamen, while caudate and globus pallidus did not reach statistical significance. This discrepancy may be due to iron accumulation slowing down earlier in globus pallidus and caudate than putamen[44]. While some QSM studies have reported significant iron accumulation in globus pallidus and caudate[57,58], no significance was reported in other studies on older populations[59,60], potentially due to the difference in the age ranges of subjects. In some subcortical ROIs, $|\chi_{dia}|$ shows increasing trends with age, which are also reported in Lao et al.[30]. The origin of these trends is unclear with no literature reporting myelin changes in these regions. In a DTI study, fractional anisotropy was reported to increase with age in putamen[61], while another study found no significant change[62]. Further exploration is necessary to understand the source of this change.

In this study, many white matter ROIs revealed decreasing trends in $|\chi_{dia}|$ over age. For example, $|\chi_{dia}|$ decreased by 1.6%/decade in the whole white matter ROI. This finding is similar to age-dependent decrease in the MWF value in the whole white matter, which was approximately 1.4%/decade (calculated by the data points manually extracted between the ages of 30 and 79 from the work by Dvorak et al.[29]). The negative linear trends of MWF with age have been demonstrated in other studies[48-50] although quadratic trends were also suggested[29,49,63], which may be attributed to the white matter maturation in early adulthood[56]. The difference in our trend may be from our older study population than the previous studies. In a few white matter ROIs, decreasing trends in $\chi_{para}$ were observed. Similarly, Lao et al.[30] reported decreasing $\chi_{para}$ after age of 40 in retrolenticular part of internal capsule. However, a postmortem iron assay study reported iron accumulation in cerebral white matter over lifespan[64], suggesting a need for further research to confirm the observation.

In contrast to a previous atlas study[30] utilizing APART-QSM technique[23], our study employed the χ-separation technique[13] to generate the atlas. The two methods reported somewhat different ROI values: In particular, substantia nigra shows $\chi_{para}$ = 115.6 ppb and $\chi_{dia}$ = –11.1 ppb at the age of 60 in our study, whereas Lao et al.[30] reported approximately $\chi_{para}$ = 79 ppb and $\chi_{dia}$ = –14 ppb (based on the manually extracted data points from the regression lines at the age of 60). Considering the QSM value of substantia nigra from a meta-analysis (112.9 ± 37.7 ppb)[65], the sum of our $\chi_{para}$ and $\chi_{dia}$ values (= 104.5 ppb) show better agreement with the meta-analysis result than that of the previous study (= 65 ppb). Similar differences are also observed in other subcortical nuclei including red nucleus and globus pallidus. In addition to these differences, we aimed to conduct a comprehensive analysis and provide quantitative statistics for 26 ROIs from MuSus-100 and ICBM-DTI-81, covering a broader scope. The proposed atlas was further analyzed for iron and myelin distributions using staining images (Fig. 4), and $\chi_{para}$ correlated with iron content (Fig. 6) and $\chi_{dia}$ correlated with MWF (Fig. 7), further consolidating the relationship between $\chi_{para}$ vs. iron and $\chi_{dia}$ vs. myelin in a large population.

In a few subjects, large calcification was observed particularly in globus pallidus, reporting high $|\chi_{dia}|$ and $\chi_{para}$ potentially due to mineral deposition of diamagnetic calcium and paramagnetic iron[66]. To avoid complications in the atlas, we excluded nine subjects with large calcification. Nevertheless, there were more subjects with small calcification that were not excluded, potentially introducing bias into the atlas. To mitigate this effect in the analysis, the median value was utilized instead of the mean value in the ROI analysis.

For the white matter ROIs, we utilized ICBM-DTI-81 atlas[42], which is based on diffusion tensor imaging (DTI). However, we noticed that the $\chi_{dia}$ values exhibited pronounced heterogeneity within ROIs, suggesting that new ROIs need to be defined through the incorporation of DTI and myelin contents.

Beyond its application in research, our atlas may be utilized in intervention treatments, specifically in DBS and HIFU. Our atlas helps accurate localization of commonly stimulated targets in DBS, such as the internal globus pallidus and subthalamic nucleus[67]. Similarly, for HIFU application, our atlas could facilitate the precise targeting of deep brain structures in the thalamus and basal ganglia[68], potentially improving the outcome of ablation.

**5. Conclusion**

In conclusion, our study successfully created a χ-separation atlas for healthy population, providing normative distributions of $\chi_{para}$ and $\chi_{dia}$ in the brain. The atlas offers insights into the spatial distributions of iron and myelin and allows us to delineate fine structures relevant to the susceptibility distributions. We expect that future research built upon our atlas will help to understand the spatial distributions of iron and myelin in the healthy human brain and their alterations in neurological disorders.

**Acknowledgements**

This research was supported by the National Research Foundation of Korea (2021R1A2B5B03002783, 2021M3E5D2A01024795, 2021R1I1A1A01040374, and 2022R1A4A1030579), INMC and IOER at Seoul National University and Heuron Co., Ltd. The ICBM-DTI-81 atlas was obtained as a courtesy of Dr. Mori (www.mristudio.org) in Johns Hopkins University. We thank for the kind permission to reproduce figures from other publications. The iron staining images in Fig. 4 were published in Naidich *et al.*, Imaging of the Brain, pp. 174-204, Copyright Elsevier (2013). The myelin staining images in Fig. 4 were published in Schaltenbrand & Wahren, Atlas for Stereotaxy of the Human Brain, Plates 51-55, Copyright Thieme (1977).

**Data availability statement**

The data and materials that support the findings and the conclusion of this study are made publicly available (https://github.com/SNU-LIST/chi-separation-atlas).

Supplementary information for

# A human brain atlas of $\chi$-separation for normative iron and myelin distributions

Kyeongseon Min, Beomseok Sohn, Woo Jung Kim, Chae Jung Park, Soohwa Song, Dong Hoon Shin, Kyung Won Chang, Na-Young Shin, Minjun Kim, Hyeong-Geol Shin, Phil Hyu Lee, Jongho Lee[*]

**Table of contents**

**Supplementary methods**

*Recruitment criteria for healthy volunteers*

      80 healthy volunteers from Yonsei Severance Hospital (YSSH) and 12 from Yongin Severance Hospital (YGSH) were recruited based on the following criteria: individuals with no history of major neurological disorders or mental disorders, no family history of movement disorders, a cross-cultural smell identification test score of 8 or higher, and a mini-mental state examination score of 27 or higher. Additionally, 24 healthy volunteers were recruited from YGSH with the following criteria: individuals with no history of major neurological disorders or mental disorders, no evidence of severe or unstable physical conditions, no history of substance use disorders within the last 3 years, individuals determined to be normal in an activities of daily living evaluation, individuals determined to be normal in a subjective cognitive decline survey, and individuals who showed normal performance in all cognitive areas in a neuropsychological test.

*Magnetization-prepared rapid gradient echo (MPRAGE) scan parameters*

      $T_1$-weighted images were acquired from 116 subjects using 3D MPRAGE sequences. Scan parameters varied as follows: TR = 9.8 ms, TE = 4.6 ms, flip angle = 8° for 80 subjects from YSSH; TR = 7.2 ms, TE = 3.3 ms, flip angle = 9° for 12 subjects from YGSH; and TR = 4.6 ms, TE = 2.0 ms, flip angle = 8° for 24 subjects from YGSH. All had a resolution of 1×1×1 mm³ and an FOV of 240×190×240 mm³. The FOV was adjusted to reduce scan time.

**Supplementary figures**

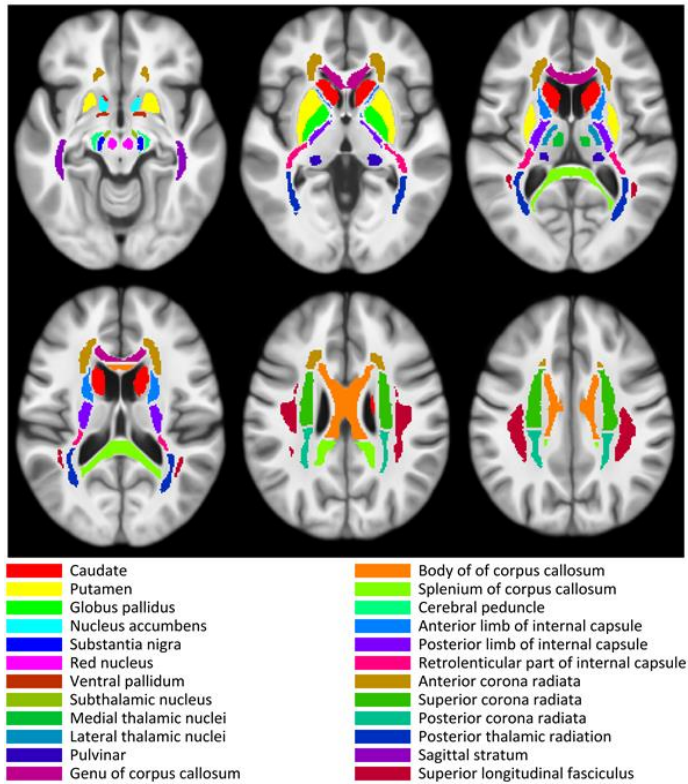

**Fig. S1.** The regions of interests (ROI) used for the analysis are overlaid on $T_1$-weighted axial slices. The corresponding color codes are presented below. The ROI labels are publicly available (https://github.com/SNU-LIST/chi-separation-atlas).

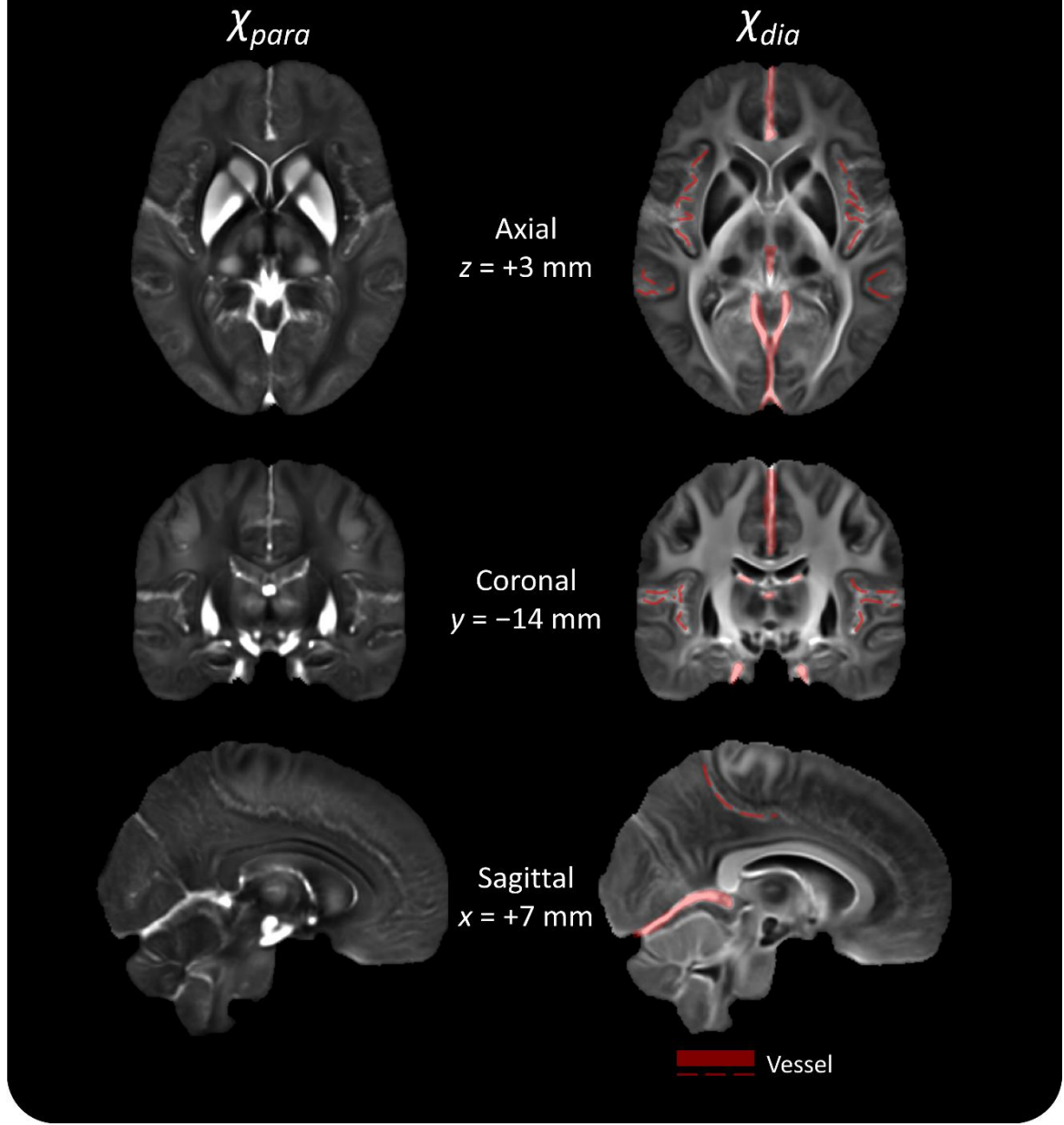

**Fig. S2.** Artifacts in χ-separation atlas. Vessel artifacts in the $\chi_{dia}$ atlas are marked with red overlay.

# Supplementary tables

**Table S1.** Population means and standard deviations of $\chi_{para}$ and $\chi_{dia}$ values in subcortical nuclei, thalamic nuclei, and white matter ROIs.

|  | ROI name | $\chi_{para}$ (ppb) | $\chi_{dia}$ (ppb) |
|---|---|---|---|
| Subcortical nuclei | Caudate | 47.7 ± 7.4 | −8.2 ± 4.9 |
|  | Putamen | 77.1 ± 20.6 | −10.3 ± 5.9 |
|  | Globus pallidus | 131.9 ± 10.4 | −12.6 ± 7.3 |
|  | Nucleus accumbens | 45.4 ± 13.2 | −23.4 ± 5.6 |
|  | Substantia nigra | 115.7 ± 14.4 | −11.1 ± 4.5 |
|  | Red nucleus | 112.0 ± 14.8 | −10.1 ± 4.1 |
|  | Ventral pallidum | 144.3 ± 27.0 | −22.1 ± 13.1 |
|  | Subthalamic nucleus | 112.1 ± 12.1 | −15.3 ± 5.9 |
| Thalamus | Medial thalamic nuclei | 31.2 ± 7.6 | −11.8 ± 5.9 |
|  | Lateral thalamic nuclei | 22.0 ± 5.2 | −22.4 ± 4.7 |
|  | Pulvinar | 52.1 ± 11.6 | −4.4 ± 3.2 |
|  | Whole thalamus | 22.9 ± 2.1 | −20.8 ± 2.5 |
| White matter | Genu of corpus callosum | 17.8 ± 1.9 | −30.8 ± 2.9 |
|  | Body of corpus callosum | 14.8 ± 1.9 | −35.1 ± 2.6 |
|  | Splenium of corpus callosum | 16.5 ± 2.6 | −42.3 ± 3.1 |
|  | Cerebral peduncle | 23.5 ± 2.8 | −43.7 ± 4.1 |
|  | Anterior limb of internal capsule | 14.5 ± 2.2 | −41.3 ± 4.0 |
|  | Posterior limb of internal capsule | 8.2 ± 2.7 | −52.2 ± 3.0 |
|  | Retrolenticular part of internal capsule | 18.0 ± 2.9 | −39.3 ± 3.3 |
|  | Anterior corona radiata | 13.9 ± 1.7 | −27.1 ± 2.9 |
|  | Superior corona radiata | 14.1 ± 1.6 | −31.7 ± 2.9 |
|  | Posterior corona radiata | 14.0 ± 2.3 | −32.0 ± 2.5 |
|  | Posterior thalamic radiation | 12.9 ± 2.5 | −41.1 ± 4.0 |
|  | Sagittal stratum | 17.3 ± 2.2 | −36.5 ± 3.9 |
|  | Superior longitudinal fasciculus | 22.6 ± 1.6 | −32.7 ± 2.9 |
|  | Whole white matter | 16.8 ± 1.1 | −28.8 ± 1.9 |

**Table S2.** Age-dependence of $\chi_{para}$ and $\chi_{dia}$ values in subcortical nuclei, thalamic nuclei, and white matter ROIs. The $\chi_{para}$ and $\chi_{dia}$ values were regressed with a linear function of age. Regression coefficients with statistical significance are marked with asterisks (*, $p < 0.05$, Bonferroni-corrected).

$$\chi_{para} \text{ or } |\chi_{dia}| \text{ (ppb)} = [\text{Slope (ppb/years)}] \times (\text{age} - 60) \text{ (years)} + [\text{Intercept (ppb)}].$$

|  | ROI name | $\chi_{para}$ (ppb) | | $|\chi_{dia}|$ (ppb) | |
|---|---|---|---|---|---|
|  |  | Slope | Intercept | Slope | Intercept |
| Subcortical nuclei | Caudate | 0.08 | 47.59* | 0.19* | 8.05* |
|  | Putamen | 0.85* | 76.40* | 0.15* | 10.16* |
|  | Globus pallidus | 0.08 | 131.84* | 0.21* | 12.43* |
|  | Nucleus accumbens | 0.59* | 44.87* | 0.15* | 23.25* |
|  | Substantia nigra | 0.07 | 115.62* | 0.01 | 11.13* |
|  | Red nucleus | 0.07 | 111.91* | 0.00 | 10.13* |
|  | Ventral pallidum | 0.38 | 143.96* | 0.32* | 21.87* |
|  | Subthalamic nucleus | 0.04 | 112.04* | 0.06 | 15.21* |
| Thalamus | Medial thalamic nuclei | –0.20* | 31.37* | 0.05 | 11.79* |
|  | Lateral thalamic nuclei | –0.17* | 22.12* | –0.02 | 22.40* |
|  | Pulvinar | –0.10 | 52.21* | –0.01 | 4.41* |
|  | Whole thalamus | –0.14* | 22.98* | 0.00 | 20.78* |
| White matter | Genu of corpus callosum | 0.03 | 17.80* | 0.01 | 30.79* |
|  | Body of corpus callosum | –0.05* | 14.51* | 0.00 | 35.10* |
|  | Splenium of corpus callosum | –0.04 | 16.54* | –0.09* | 42.40* |
|  | Cerebral peduncle | –0.04 | 23.51* | –0.01 | 43.68* |
|  | Anterior limb of internal capsule | 0.01 | 14.51* | –0.14* | 41.43* |
|  | Posterior limb of internal capsule | –0.07* | 8.29* | –0.08* | 52.23* |
|  | Retrolenticular part of internal capsule | –0.06 | 18.07* | –0.12* | 39.45* |
|  | Anterior corona radiata | –0.01 | 13.90* | –0.10* | 27.18* |
|  | Superior corona radiata | –0.04* | 14.17* | –0.07* | 31.72* |
|  | Posterior corona radiata | –0.08* | 14.03* | –0.03 | 32.01* |
|  | Posterior thalamic radiation | –0.10* | 12.95* | –0.05 | 41.09* |
|  | Sagittal stratum | –0.01 | 17.26* | –0.05 | 36.54* |
|  | Superior longitudinal fasciculus | –0.01 | 22.56* | –0.06* | 32.72* |
|  | Whole white matter | –0.02 | 16.82* | –0.05* | 28.81* |